\documentclass[reprint,amsmath,amssymb,aps,pre,longbibliography]{revtex4-1}

\usepackage[dvips]{graphicx}
\usepackage{color}
\usepackage{pstricks,pst-grad}
\usepackage{amsbsy}
\usepackage{epic}
\usepackage[normalem]{ulem}
\usepackage{bm}

\usepackage{hyperref}
\usepackage[hyphenbreaks]{breakurl}

\newcommand\beq{\begin{equation}}
\newcommand\be{\begin{equation}}
\newcommand\ee{\end{equation}}
\newcommand\eeq{\end{equation}}
\newcommand\beqa{\begin{eqnarray}}
\newcommand\eeqa{\end{eqnarray}}
\newcommand{\nn}{\nonumber\\}
\newcommand\ba{\begin{eqnarray}}
\newcommand\ea{\end{eqnarray}}
\newcommand\ocite{\onlinecite}

\def\bal#1\eal{\begin{align}#1\end{align}}

\newcommand{\longwidth}{1.8\columnwidth}

\newcommand{\e}{\eta}
\newcommand{\la}{\ell}

\newcommand{\Aa}{a}
\newcommand{\Bb}{b}
\newcommand{\Cc}{c}
\newcommand{\Ss}{\mathcal{S}}

\begin{document}

\title{Structural properties of the Jagla fluid}

\author{Mariano L\'{o}pez de Haro}
\email{malopez@unam.mx}
\homepage{http://xml.cie.unam.mx/xml/tc/ft/mlh/}
\affiliation{Instituto de Energ\'{\i}as Renovables, Universidad Nacional Aut\'onoma de M\'exico (U.N.A.M.),
Temixco, Morelos 62580, M{e}xico}
\author{\'Alvaro Rodr\'{\i}guez-Rivas}
\email{arodriguezrivas@unex.es}
\homepage{https://arodriguez-rivas.weebly.com/}
\author{Santos B. Yuste}
\email{santos@unex.es}
\homepage{http://www.unex.es/eweb/fisteor/santos/}
\author{Andr\'es Santos}
\email{andres@unex.es}
\homepage{http://www.unex.es/eweb/fisteor/andres/}
\affiliation{Departamento de F\'{\i}sica  and Instituto de Computaci\'on Cient\'{\i}fica Avanzada (ICCAEx), Universidad de
Extremadura, Badajoz, E-06006, Spain}

\begin{abstract}
The structural properties of the Jagla fluid are studied by Monte Carlo (MC) simulations, numerical solutions of integral equation theories, and  the (semi-analytical) rational-function approximation (RFA) method. In the latter case, the results are obtained from the  assumption (supported by our MC simulations) that the Jagla potential and a potential with a hard core plus an appropriate piecewise constant function lead to practically the same cavity function.
The predictions obtained for the radial distribution function, $g(r)$, from this approach
are compared against MC simulations and  integral equations for the Jagla model, and also for the limiting cases of the triangle-well potential and the ramp potential, with a general good agreement. The analytical form of the RFA in Laplace space allows us to describe the asymptotic behavior of $g(r)$ in a clean way and compare it with MC simulations for representative states with oscillatory or monotonic decay. The RFA predictions for the Fisher--Widom and Widom lines of the Jagla fluid are  obtained.
\end{abstract}

\date{\today}

%\keywords{}

\maketitle

\section{Introduction}
\label{intro}

{Substances such as water, phosphorous, carbon, or silicon, whose intermolecular interactions depend on the orientation between their bonds, can all exhibit anomalous static and/or dynamic behavior. In order to understand and deal with such behavior, it is convenient to rely on simple models which may shed light on the mechanisms leading to it. Ever since the appearance of the  pioneering work by Hemmer and Stell \cite{HS70,SH72} on fluid systems whose molecules interact through spherically symmetric core-softened pair potentials,  these and closely related models have been considered in the literature for the previous purpose. Among the problems that have been addressed, one finds such models in attempts to describe reentrant melting \cite{DRB91,MSP09, PSM10, LMBS10, MPS11,MS11}, thermodynamic anomalies \cite{DRB91,J98b,SSBS98,J99a, SSGBS00, SSGBS01, WM02, J04, YBGS05, KBSZS05, ONCB06, YBGDS06, XBAS06, CP06, F07, OFNB08, ONB08, ONB09,BMAGPSSX09, GFFR09, PDDS09, SPM09, FS10,PSS11,VF11,BSB11,NC13,CS13,HD13,HU13,HU14,Y15, MS16,SXG17,LHP17,HKAK18}, anomalous transport \cite{ NRCB04, XEBS06, XMYSBS09, FTR11b, FTR13, NC13,  BF15, SXG17,HKAK18}, liquid-liquid phase transitions and phase diagrams \cite{ SAEHPS94,VMNHS00, J01a,J01b, FMSBS01, MP01, BFGMSSSS02b, RS03, SBFMS04, XKBCPSS05, CP06, GW06, ZJWS07, SKXYMBCM07,BCT07, LAMM07,YBKGS08, HLA09, XBGS10, BSB11,HD13,HMU14, LXLSB14, LXASB15, MU15, CJBTS15, RD17, LHP17, ADCARS18},  and glassy behavior \cite{J01b,KBSZS05, XBAS06, BMAGPSSX09, PDDS09, XBGS10,BF15, LXASB15,J99a, ABG11, A15a, A15b}.}

\begin{figure}
\includegraphics[width=.9\columnwidth]{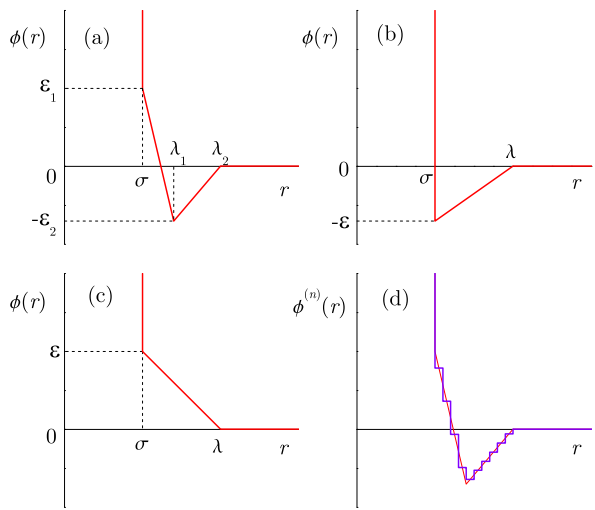}
\caption{Sketch of (a) the Jagla potential, (b) the triangle-well potential, and (c) the ramp potential. (d) A sketch of the discretized version of the Jagla potential with $n=10$.
\label{fig:sketch}}
\end{figure}

 {Among the family of core-softened pair potentials, one that is able to predict multiple fluid transitions and some of the water-type thermodynamic and dynamic anomalies is the Jagla ramp potential (hard core plus a linear repulsive ramp and a linear attractive ramp) \cite{J99a} given by}
    \begin{equation}
\label{Jagla}
\phi(r)=
\begin{cases}
\infty  ,& r<\sigma, \\
\displaystyle{\frac{\varepsilon_1(\lambda_1-r)-\varepsilon_2(r-\sigma)}{\lambda_1-\sigma}} ,& \sigma\leq r\leq\lambda_1, \\
\displaystyle{-\frac{\varepsilon_2(\lambda_2-r)}{\lambda_2-\lambda_1}} ,& \lambda_1 \leq r\leq \lambda_2, \\
0,&r\geq\lambda_2.
\end{cases}
\end{equation}
The explicit form of this potential is determined by five parameters, namely three lengths (the hard-core diameter $\sigma$ and the ranges $\lambda_1$ and $\lambda_2$) and two energies (the height $\varepsilon_1$ of the repulsive ramp and the depth $\varepsilon_2$ of the attractive well, both taken to be positive). It includes the two interesting limiting cases of the triangle-well potential ($\lambda_1=\sigma$, $\lambda_2=\lambda$, $\varepsilon_2=\varepsilon$) and the ramp potential ($\varepsilon_2=0$, $\varepsilon_1=\varepsilon$, $\lambda_1=\lambda_2=\lambda$).
Figures \ref{fig:sketch}(a)--\ref{fig:sketch}(c) sketch the Jagla, triangle-well, and ramp potentials, respectively.

For such fluids,  we  take $\sigma = 1$ as the length unit and $\varepsilon_\text{ref}=\varepsilon_2 = 1$ (Jagla potential), $\varepsilon_\text{ref}=\varepsilon=1$ (triangle-well potential), or $\varepsilon_\text{ref}=\varepsilon=1$ (ramp potential) as the energy unit.
Reduced units ($\rho^{*}=\rho\sigma^{3}$ for the density and $T^{*}=k_{B}T/\varepsilon_\text{ref}$ for the temperature) are used, so that the additional relevant parameters are $\varepsilon_1/\varepsilon_2$, $\lambda_1$, and $\lambda_2$ for the Jagla potential, and $\lambda$ for the triangle-well and ramp potentials.

{While there has been a lot of work in the literature concerning the thermodynamic properties of the Jagla fluid, as far as we know no work other than the paper by Gibson and Wilding \cite{GW06} has been devoted to the structural properties of a fluid whose molecules interact through such a potential.} Therefore, the major aim of this paper is to present a semi-analytical approach based on the rational-function approximation (RFA) \cite{SYH12,SYHBO13,S16} to obtain the radial distribution function $g(r)$ of the Jagla fluid, including its asymptotic behavior for large $r$.
The application of the RFA to the Jagla fluid is made by assuming that a representation of the potential consisting in a hard core plus an appropriate piecewise constant function leads to essentially the same cavity function as the original Jagla potential.
The outcome of the RFA approach will be assessed by testing its validity against integral equation results [both the Percus--Yevick (PY) and hypernetted-chain  (HNC) approximations will be considered] and our own Monte Carlo (MC) simulation data.

The paper is organized as follows. Section \ref{sect1b} discusses how the radial distribution function of the Jagla fluid can be approximately obtained from that of a discretized $n$-step potential. Next, in order to make the paper self-contained, in Sec.\ \ref{sect2} we summarize the RFA method leading to the structural properties of the $n$-step fluid.
This is followed in Sec.\ \ref{sect3} by the comparison between the results obtained with the present approach for various cases of Jagla, triangle-well, and ramp fluids, and those of the PY and HNC approximations when tested against our MC simulation data.
Section \ref{sect4} deals with the Fisher--Widom and Widom lines, as well as with the static structure factor, of the Jagla fluid.
The paper is closed in Sec.\ \ref{sec:concl} with some discussion and concluding remarks.

\section{Mapping to piecewise constant potentials}
\label{sect1b}

Suppose we want to replace the actual potential \eqref{Jagla} by a \emph{discretized} version $\phi^{(n)}(r)$ consisting of a sequence of $n$ steps of ``heights'' $\epsilon_j$ and widths $(\la_j-\la_{j-1})$ (with the conventions $\la_0=\sigma$ and $\la_n=\lambda_2$), namely \cite{BG99,LXASB15}
     \begin{equation}
\label{a1}
\phi^{(n)}(r)=
\begin{cases}
\infty  ,& r<\sigma, \\
\epsilon_1  ,& \sigma<r<\la_1 , \\
\epsilon_2  ,& \la_1<r<\la_2 , \\
\vdots&\vdots \\
\epsilon_n  ,& \la_{n-1}<r<\la_n , \\
0,&r>\la_n.
\end{cases}
\end{equation}
The simplest choice for $\phi^{(n)}(r)$ to mimic the original potential $\phi(r)$ is to take $\epsilon_j=\phi((\ell_{j-1}+\ell_j)/2)$ and a constant step width $\Delta r=(\lambda_2-\sigma)/n$, so that $\ell_j=\sigma+j \Delta r$.
This is illustrated in Fig.\ \ref{fig:sketch}(d) for the case $n=10$.

Once the representation $\phi(r)\to \phi^{(n)}(r)$ has been done, the next step in the mapping consists in approximating the cavity function,
\beq
y(r)\equiv g(r)e^{\beta\phi(r)},\quad \beta\equiv\frac{1}{k_BT},
 \eeq
of the Jagla fluid  with the one of the $n$-step fluid, $ y^{(n)}(r)\equiv g^{(n)}(r)e^{\beta \phi^{(n)}(r)}$, namely
\beq
\label{yJagla}
y(r)\approx y^{(n)}(r)
\eeq
or, equivalently,
\begin{equation}
\label{gJagla}
g(r)\approx g^{(n)}(r)e^{\beta\left[\phi^{(n)}(r)-\phi(r)\right]}.
\end{equation}
Clearly, the practical usefulness of the method rests on considering a relatively small number of steps in the discretization.

\begin{figure}
\includegraphics[width=.9\columnwidth]{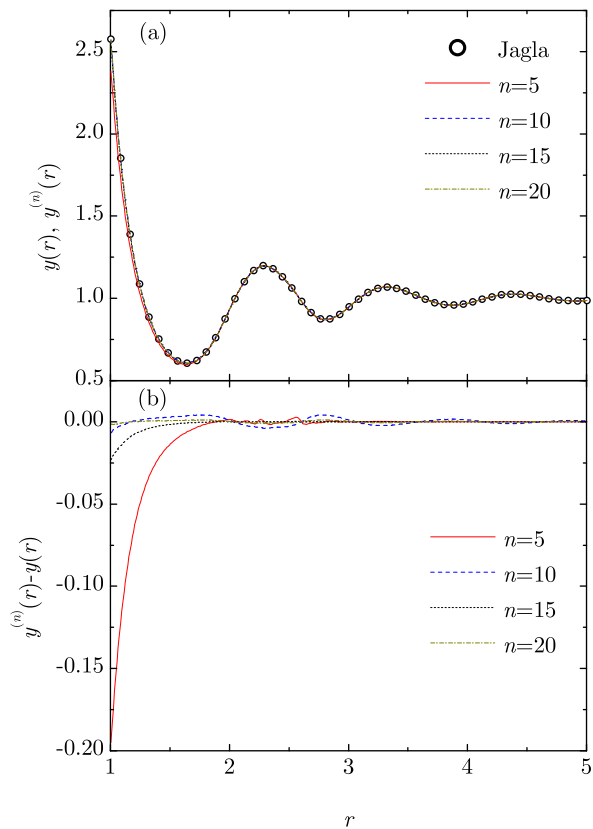}
\caption{(a) MC results for the cavity function at $\rho^*=0.6$  and $T^*=1$. The circles correspond to the true Jagla potential with $\lambda_1=1.3$, $\lambda_2=1.6$, and $\varepsilon_1/\varepsilon_2=1$, while the lines correspond to the discretized version of the same potential with $n=5$ (---), $n=10$ (- - -), $n=15$ ($\cdots$), and $n=20$ (--$\cdot$--$\cdot$--). (b) Difference between the MC cavity function of the discretized potential and the MC cavity function of the genuine Jagla potential.
\label{fig:Steps}}
\end{figure}

\begin{figure}
\includegraphics[width=.9\columnwidth]{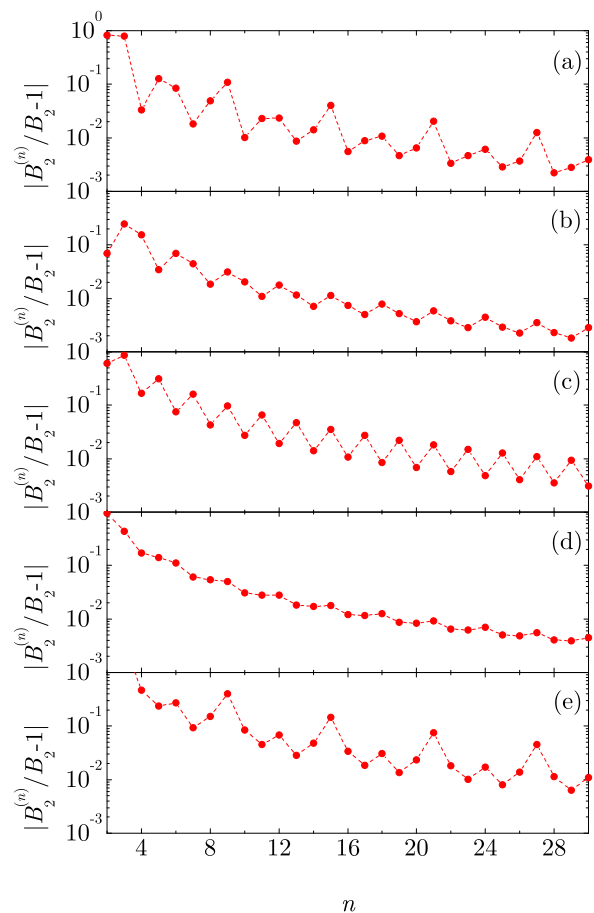}
\caption{Relative difference $\left|{B_2^{(n)}}/{B_2}-1\right|$ versus the number of steps $n$ for the Jagla potential with $\varepsilon_1/\varepsilon_2=1$, $T^*=1$, $\lambda_2=1.6$, and  (a) $\lambda_1=1.1$, (b) $\lambda_1=1.2$, (c) $\lambda_1=1.3$, (d) $\lambda_1=1.4$, and (e) $\lambda_1=1.5$. The lines are a guide to the eye.
\label{fig:B2}}
\end{figure}

While the MC results that are  presented in Secs.\ \ref{sect3} and \ref{sect4} have been obtained for the true Jagla potential \eqref{Jagla}, we have performed additional MC simulations for the discretized potential \eqref{a1}, with different choices of $n$, in order to test the reliability of the ansatz \eqref{yJagla}.

{For illustration purposes, in Figs.\ \ref{fig:Steps}(a) and \ref{fig:Steps}(b) we show, respectively, the cavity functions $y(r)$ and $y^{(n)}(r)$,  and the deviations $y^{(n)}(r)-y(r)$ (with $n=5$, $n=10$, $n=15$, and $n=20$) for a representative  Jagla fluid. Here, the potential parameters correspond to $\lambda_1=1.3$, $\lambda_2=1.6$, and $\varepsilon_2=\varepsilon_1$,  while the reduced density and temperature are $\rho^*=0.6$  and $T^*=1$, respectively.
One can observe in Fig.\ \ref{fig:Steps} that the functions $y^{(n)}(r)$ are practically indistinguishable from the true Jagla cavity function $y(r)$. It is also clear that the cavity function $y^{(n)}(r)$ with $n =5$ underestimates the contact value and the values up to the first minimum, while such  limitations are widely corrected with $n = 10$, $n=15$ and, especially, $n = 20$. To the best of our knowledge, the strong insensitivity of the cavity function to the ``details'' of the potential (such as the number of steps), and hence the practical validity of Eqs.\ \eqref{yJagla} and \eqref{gJagla}, has not been noted before.

Nevertheless, a word of caution and some technical issues are pertinent here, since not everything is as clearcut as the previous analysis could suggest. Paradoxically, the differences $y^{(n)}(r)-y(r)$ with $n=10$ are typically smaller than with $n =15$. This shows that the convergence $y^{(n)}(r)\to y(r)$ is not monotonic with increasing $n$. In particular, we have observed that in the case $(\lambda_1,\lambda_2)=(1.3,1.6)$ the \emph{optimal} values are $n=\text{even}$, but other choices are more appropriate for other values of $(\lambda_1,\lambda_2)$.

A convenient predictor of the optimal values of $n$ relies on the use of the second virial coefficient \cite{HM06,S16}
\begin{equation}
  B_2=-2\pi\int_0^\infty dr\, r^2 \left[e^{-\beta\phi(r)}-1\right].
\end{equation}
Our MC simulations suggest that a  reliable criterion to predict the qualitative dependence on $n$  of the difference between $y^{(n)}(r)$ and $y(r)$  is based on the analysis of the relative difference $\delta B_2^{(n)} \equiv\left|{B_2^{(n)}}/{B_2}-1\right|$  between the second virial coefficient, $B_2^{(n)}$, of the discretized potential $\phi^{(n)}(r)$ and the second virial coefficient, $B_2$, of the continuous potential $\phi(r)$.  Figure \ref{fig:B2} shows that $\delta B_2^{(n)}$ presents local minima at certain (optimal) values of $n$, which follow regular patterns depending on the values of $(\lambda_1,\lambda_2)$.
For instance, in the case of Fig.\ \ref{fig:B2}(e), one can observe that the choice $n=13$ is better than the higher values $n=14$, $15$, and $16$.
We have checked that the qualitative dependence of $\delta B_2^{(n)}$ on $n$ is not affected by the values of $T^*$ and $\varepsilon_1/\varepsilon_2$. Finally, although not shown, it should be stressed that  the convergence with increasing $n$ is monotonic in the cases of the ramp and triangle-well potentials. In any case, the choice $n=10$ seems to be a reasonable one in most instances.}

\section{RFA approach for piecewise constant potentials with a hard core}
\label{sect2}

In this section we provide the essential steps leading to the computation of the radial distribution function $g^{(n)}(r)$ of the fluid with the $n$-step potential \eqref{a1}. For further details, we refer the reader to Refs.\ \ocite{SYH12,SYHBO13,S16}. We begin by expressing  the Laplace transform of $rg^{(n)}(r)$ as
\begin{equation}
\label{b5}
G^{(n)}(s)=s\frac{F^{(n)}(s)e^{-s}}{1+12\eta F^{(n)}(s)e^{- s}},
\end{equation}
where  $\eta \equiv \frac{\pi}{6}\rho^*$ is the packing fraction. Further, $F^{(n)}(s)$ is decomposed  as
\begin{equation}
F^{(n)}(s)=\sum_{j=0}^n R_j^{(n)}(s)e^{-(\la_j-1)s}
\label{c0}
\end{equation}
to reflect the discontinuities of $g^{(n)}(r)$ at  $r=\la_j$.

We next assume the following \emph{rational-function} approximation for $R_j^{(n)}(s)$:
\bal
R_j^{(n)}(s)=&-\frac{1}{12\eta}\frac{\Aa_j^{(n)}+\Bb_j^{(n)} s}{1+\Ss_1^{(n)} s+\Ss_2^{(n)} s^2+\Ss_3^{(n)} s^3}, \nn &\qquad j=0,\ldots,n.
\label{c6}
\eal
The $2n+5$ constants $\{\Aa_j^{(n)}, \Bb_j^{(n)};j=0,\ldots,n\}$ and $\{\Ss_k^{(n)};k=1,2,3\}$ must satisfy certain consistency conditions. First, the exact physical requirement $G^{(n)}(s)=s^{-2}+{O}(s^0)$ implies
\begin{subequations}
\label{c7-c12}
\beq
\Aa_0^{(n)}=1-\sum_{j=1}^n \Aa_j^{(n)},
\label{c7}
\eeq
\beq
\Bb_0^{(n)}=\Cc_1^{(n)}+\frac{\e/2}{1+2\e}\left[6\Cc_2^{(n)}+4\Cc_3^{(n)}+\Cc_4^{(n)}\right]+\frac{1+\e/2}{1+2\e},
\label{c11}
\eeq
\beq
\Ss_1^{(n)}=\Bb_0^{(n)}-\Cc_1^{(n)}-1,
\label{c8}
\eeq
\beq
\Ss_2^{(n)}=\frac{1}{2}-\Bb_0^{(n)}+\Cc_1^{(n)}+\frac{1}{2}\Cc_2^{(n)},
\label{c9}
\eeq
\beq
\Ss_3^{(n)}=\frac{1}{2}\Bb_0^{(n)}-\frac{1}{2}\Cc_1^{(n)}-\frac{1}{2}\Cc_2^{(n)}-\frac{1}{6}\Cc_3^{(n)}-\frac{1+2\e}{12\e},
\label{c10}
\eeq
\end{subequations}
where
\beq
\Cc_k^{(n)}\equiv \sum_{j=1}^n \left[\Aa_j^{(n)} (\la_j-1)^k-k
\Bb_j^{(n)}(\la_j-1)^{k-1}\right].
\label{c12}
\eeq
Thus, the five coefficients $\Aa_0^{(n)}$, $\Bb_0^{(n)}$,  $\Ss_1^{(n)}$, $\Ss_2^{(n)}$, and $\Ss_3^{(n)}$ are linear combinations of the $2n$ parameters $\{\Aa_j^{(n)}, \Bb_j^{(n)};j=1,\ldots,n\}$.
Next, one must account for the fact that the cavity function $y^{(n)}(r)$
is continuous at $r=\la_j$. Specifically,
\bal
\frac{\Bb_j^{(n)}}{\Ss_3^{(n)}}=&\sum_{\alpha=1}^3\frac{\left[e^{\beta(\epsilon_{j}-\epsilon_{j+1})}-1\right]s_\alpha^{(n)} e^{\la_j s_\alpha^{(n)}}}{\Ss_1^{(n)} +2\Ss_2^{(n)} s_\alpha^{(n)}+3\Ss_3^{(n)} s_\alpha^{(n)2}}\nn
&\times\sum_{i=0}^{j-1}\left[\Aa_i^{(n)}+\Bb_i^{(n)}s_\alpha^{(n)}\right]e^{-\la_i s_\alpha^{(n)}},\quad
j=1,\ldots,n,
\label{c15_c}
\eal
with the convention $\epsilon_{n+1}=0$ and where $s_\alpha^{(n)}$ ($\alpha=1,2,3$) are the three roots of the cubic equation $1+\Ss_1^{(n)}s+\Ss_2^{(n)}s^2+\Ss_3^{(n)}s^3=0$.
Finally, for simplicity,  the parameters $\{\Aa_j^{(n)}\}$ are set to their low-density values, namely
\beq
\Aa_j^{(n)}=e^{-\beta\epsilon_{j+1}}-e^{-\beta\epsilon_{j}},\quad j=1,\ldots,n.
\label{cc13j}
\eeq

Insertion of Eqs.\ \eqref{c7-c12} and \eqref{cc13j} into Eq.\ \eqref{c15_c} yields a  set of $n$ coupled transcendental equations for $\{\Bb_j^{(n)};j=1,\ldots,n\}$ that has to be solved numerically. This task can be undertaken with the help of a computer algebra system and turns out to be rather manageable, even for $n=20$. Once the solution is found, Eqs.\ \eqref{b5}--\eqref{c6} provide the \emph{explicit} $s$ dependence of the Laplace transform $G^{(n)}(s)$. Laplace inversion of Eq.\ (\ref{b5}) yields a useful representation of $g^{(n)}(r)$ as
 \begin{equation}
 \label{b6}
 g^{(n)}(r)=r^{-1}\sum_{m=1}^\infty (-12\eta)^{m-1}f_m^{(n)}(r-m)\Theta(r-m),
 \end{equation}
where $f_m^{(n)}(r)$ is the inverse Laplace transform of $s[F^{(n)}(s)]^m$ and $\Theta(r)$
is the Heaviside step function.
In particular, the contact value is
\beq
g^{(n)}(1^+)=\lim_{s\to\infty}s^2 F^{(n)}(s)=\frac{\Bb_0^{(n)}}{\Ss_3^{(n)}}.
\eeq
We have checked that the dependence of $y^{(n)}(1)$ on $n$ follows essentially the same pattern as the second virial coefficient $B_2^{(n)}$.

Once the RFA provides the radial distribution function $g^{(n)}(r)$ for the $n$-step potential \eqref{a1}, the ansatz \eqref{gJagla} allows one to obtain the RFA prediction for the continuous potential.

\begin{figure*}
\includegraphics[width=\longwidth]{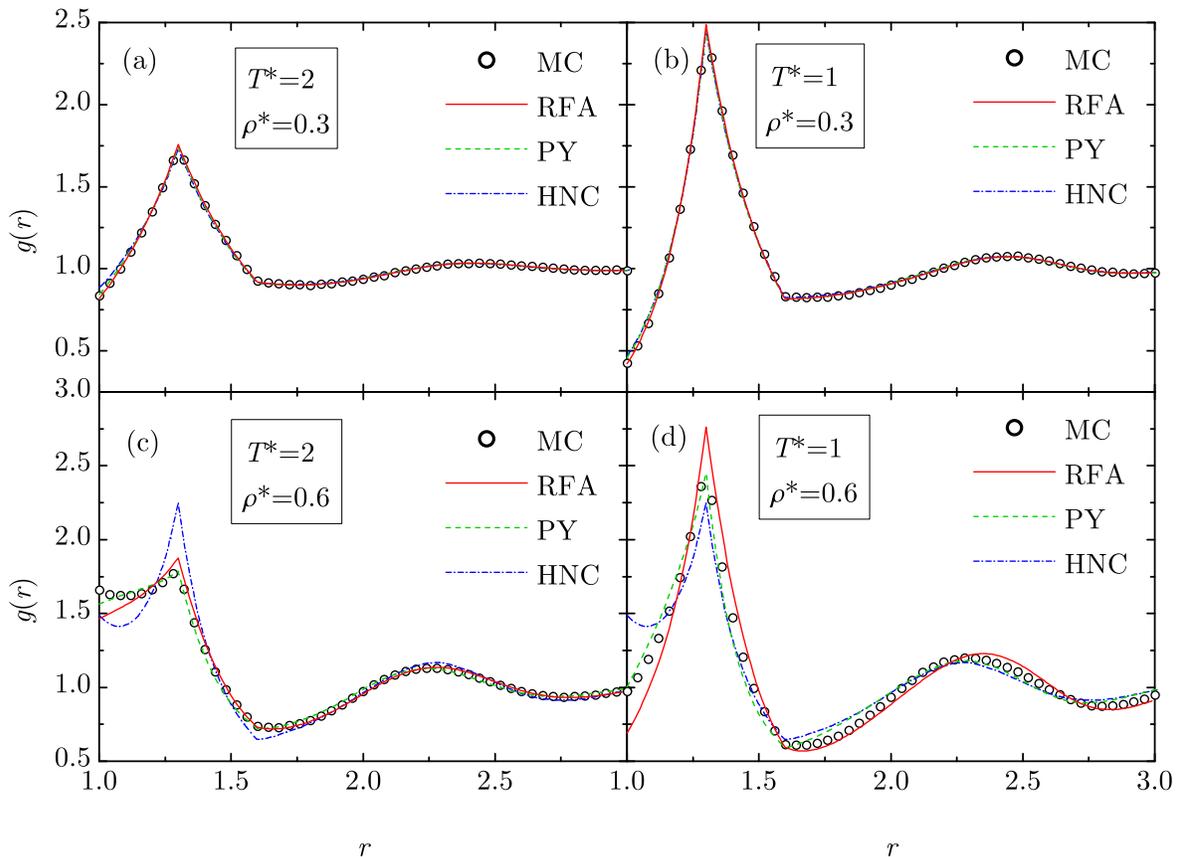}
\caption{Radial distribution function of a fluid with a Jagla potential ($\lambda_1=1.3$, $\lambda_2=1.6$, $\varepsilon_1/\varepsilon_2=1$) at (a) $(T^*,\rho^*)=(2,0.3)$, (b) $(T^*,\rho^*)=(1,0.3)$, (c) $(T^*,\rho^*)=(2,0.6)$, and (d) $(T^*,\rho^*)=(1,0.6)$. The symbols, solid lines, dashed lines, and dash-dotted lines correspond to MC simulations, the RFA with $n=10$, the PY integral equation, and the HNC integral equation, respectively.
\label{fig:Jagla}}
\end{figure*}

\begin{figure*}
\includegraphics[width=\longwidth]{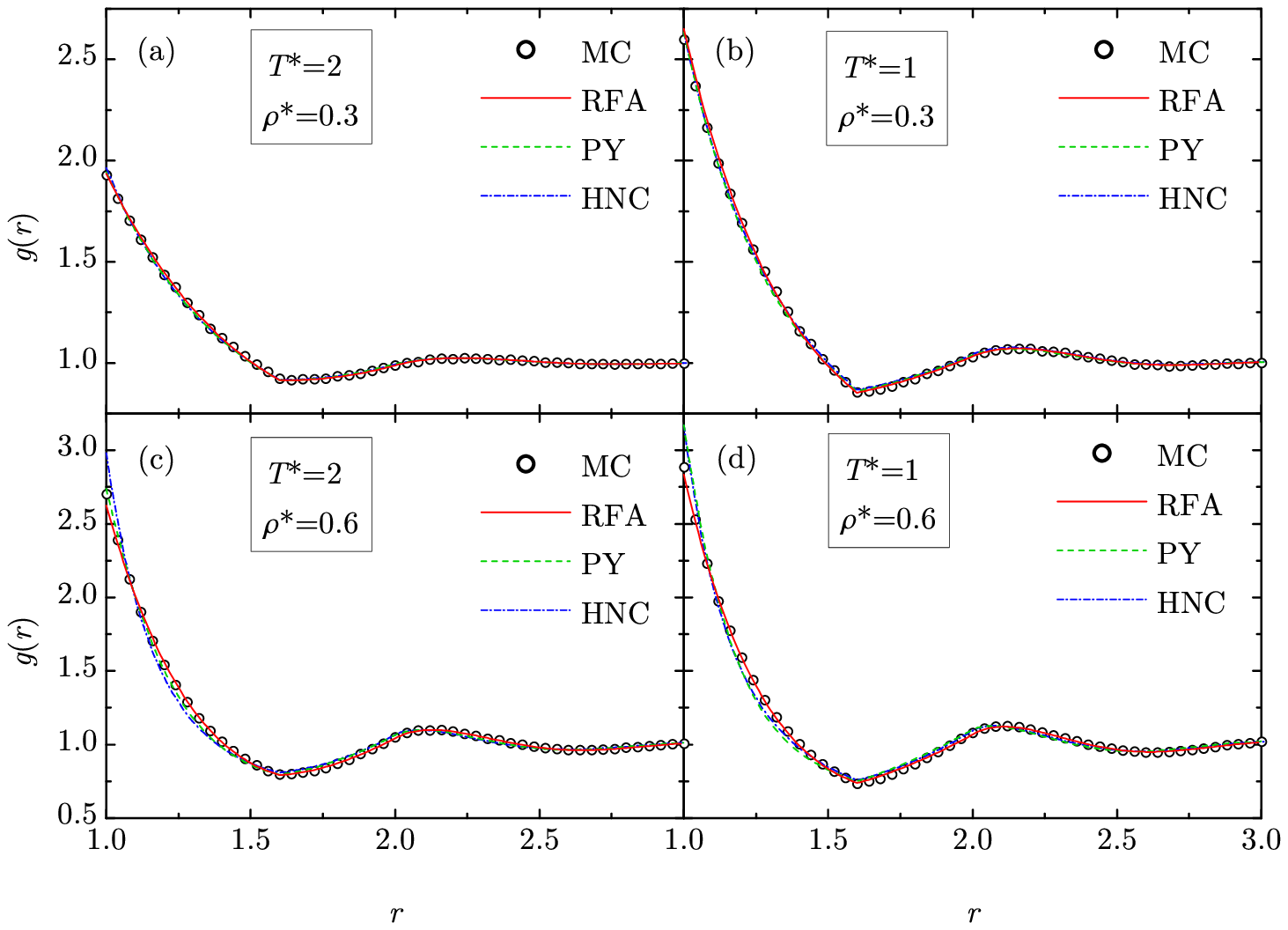}
\caption{Radial distribution function of a fluid with a triangle-well potential ($\lambda=1.6$) at (a) $(T^*,\rho^*)=(2,0.3)$, (b) $(T^*,\rho^*)=(1,0.3)$, (c) $(T^*,\rho^*)=(2,0.6)$, and (d) $(T^*,\rho^*)=(1,0.6)$. The symbols, solid lines, dashed lines, and dash-dotted lines correspond to MC simulations, the RFA with $n=10$, the PY integral equation, and the HNC integral equation, respectively.
\label{fig:Well}}
\end{figure*}

\begin{figure*}
\includegraphics[width=\longwidth]{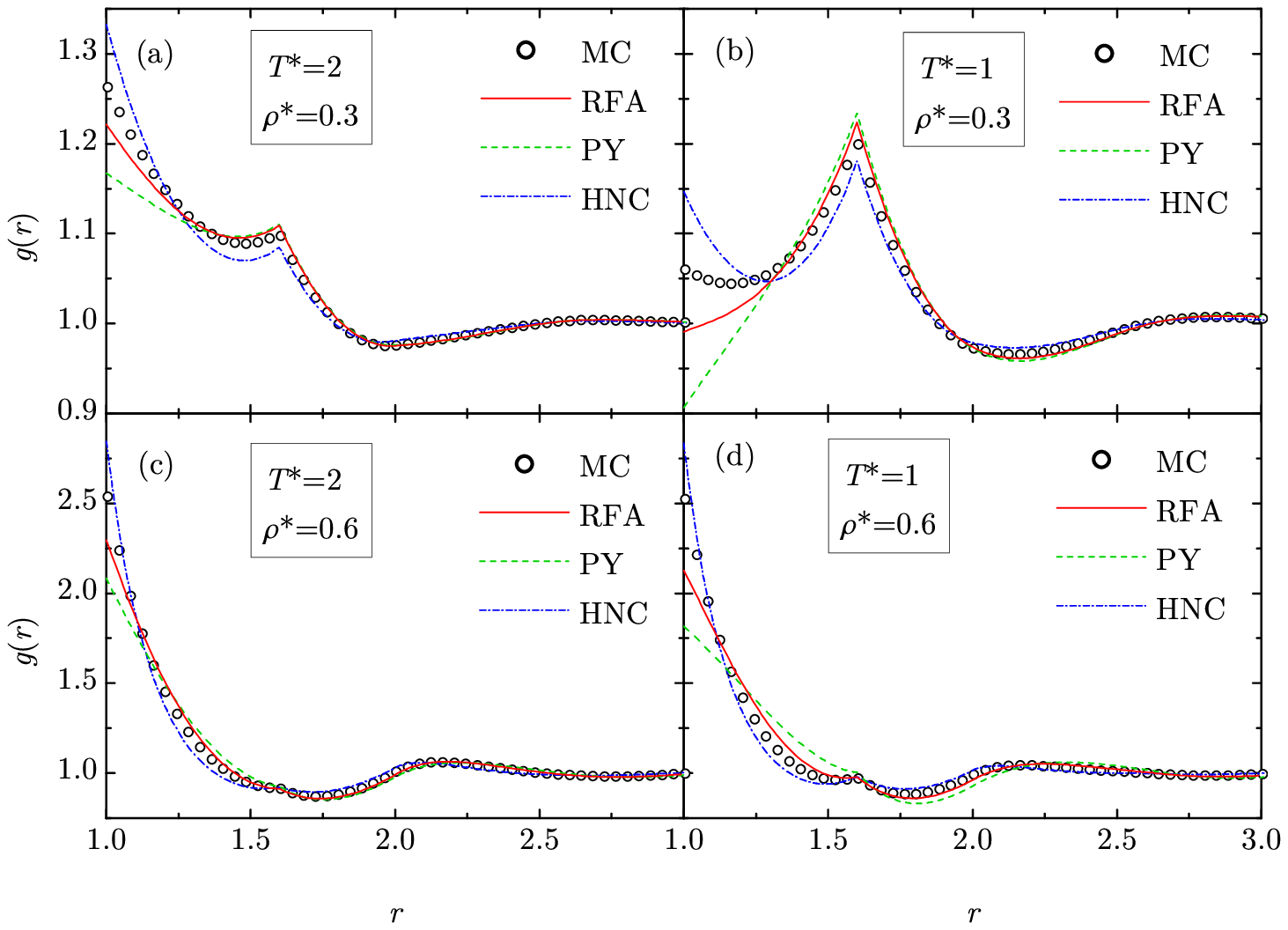}
\caption{Radial distribution function of a fluid with a ramp potential ($\lambda=1.6$) at (a) $(T^*,\rho^*)=(2,0.3)$, (b) $(T^*,\rho^*)=(1,0.3)$, (c) $(T^*,\rho^*)=(2,0.6)$, and (d) $(T^*,\rho^*)=(1,0.6)$. The symbols, solid lines, dashed lines, and dash-dotted lines correspond to MC simulations, the RFA with $n=10$, the PY integral equation, and the HNC integral equation, respectively.
\label{fig:Ramp}}
\end{figure*}

\section{Radial distribution function. MC simulations and theoretical results}
\label{sect3}

In this section we present the results of the computations performed with the RFA approach and those of NVT MC simulations of a fluid of particles that interact with the Jagla, triangle-well, and ramp potentials (see Fig.\ \ref{fig:sketch}). Numerical solutions of the PY and HNC  integral equations are also presented.
It must be remarked that, as said before, the MC simulations were performed on the true Jagla potential \eqref{Jagla}. The same happens with the numerical solutions of the PY and HNC integral equations. On the other hand, the results corresponding to the RFA were obtained by the method described in Sec.\ \ref{sect2} and application of Eq.\ \eqref{gJagla}.

{In the MC simulations, we considered a system with $N = 1372$ particles. The system was simulated during $10^6$ MC steps  (each one consisting of $N$ displacement attempts) for equilibration plus $2\times 10^6$ additional MC steps to collect data every $50$ MC steps,  taking averages every $1000$ recorded data points. Moreover, $50$ independent simulations were performed for each given case.  The radial distribution function $g(r)$ was averaged over the $50$ simulations and evaluated with a bin size $\delta r=0.01$.}

For the sake of illustration, in what follows we have taken $\lambda=1.6$ for the triangle-well and ramp potentials, and $\lambda_1=1.3$, $\lambda_2=1.6$, and $\varepsilon_1/\varepsilon_2=1$ for the Jagla potential. We have chosen four representative states ($\rho^*=0.3$ and $0.6$, $T^*=1$ and $2$).
The results are displayed in Figs.\ \ref{fig:Jagla}--\ref{fig:Ramp}.

It is worth pointing out that, for the cases presented, the RFA approach leads to very good agreement with the MC simulation results. In fact, such an agreement is as good as, and in some instances better than, the solution of either the PY or the HNC integral equations.
This is especially noteworthy since the RFA only requires the numerical solution of a few (actually $n$) nonlinear equations, whereas the PY and HNC integral equations are numerically solved by introducing a cutoff distance, replacing the integrals by sums  involving a large number of unknown local values, and iterating  until certain convergence criteria are satisfied.

As expected \cite{SYH12,SYHBO13}, the performance of the RFA tends to worsen as density increases and/or temperature decreases, especially near contact. On the other hand, even in those cases, the oscillations of $g(r)$ for larger distances are well accounted for, a feature that is addressed with more detail in Sec.\ \ref{sect4}.

Also remarkable is the fact that, with a relatively small number of steps (in all of these cases we took $n=10$), one gets a rather reasonable description of the structure of the fluid. This shows that, as confirmed by Fig.\ \ref{fig:Steps}, the replacement of the original potential $\phi(r)$ by a discretized version $\phi^{(n)}(r)$, and the subsequent application of Eq.\ \eqref{gJagla}, indeed represent a practical route to obtain the structural properties of the fluid.

\section{Fisher--Widom and Widom lines for the Jagla fluid}
\label{sect4}

\begin{figure}
\includegraphics[width=.9\columnwidth]{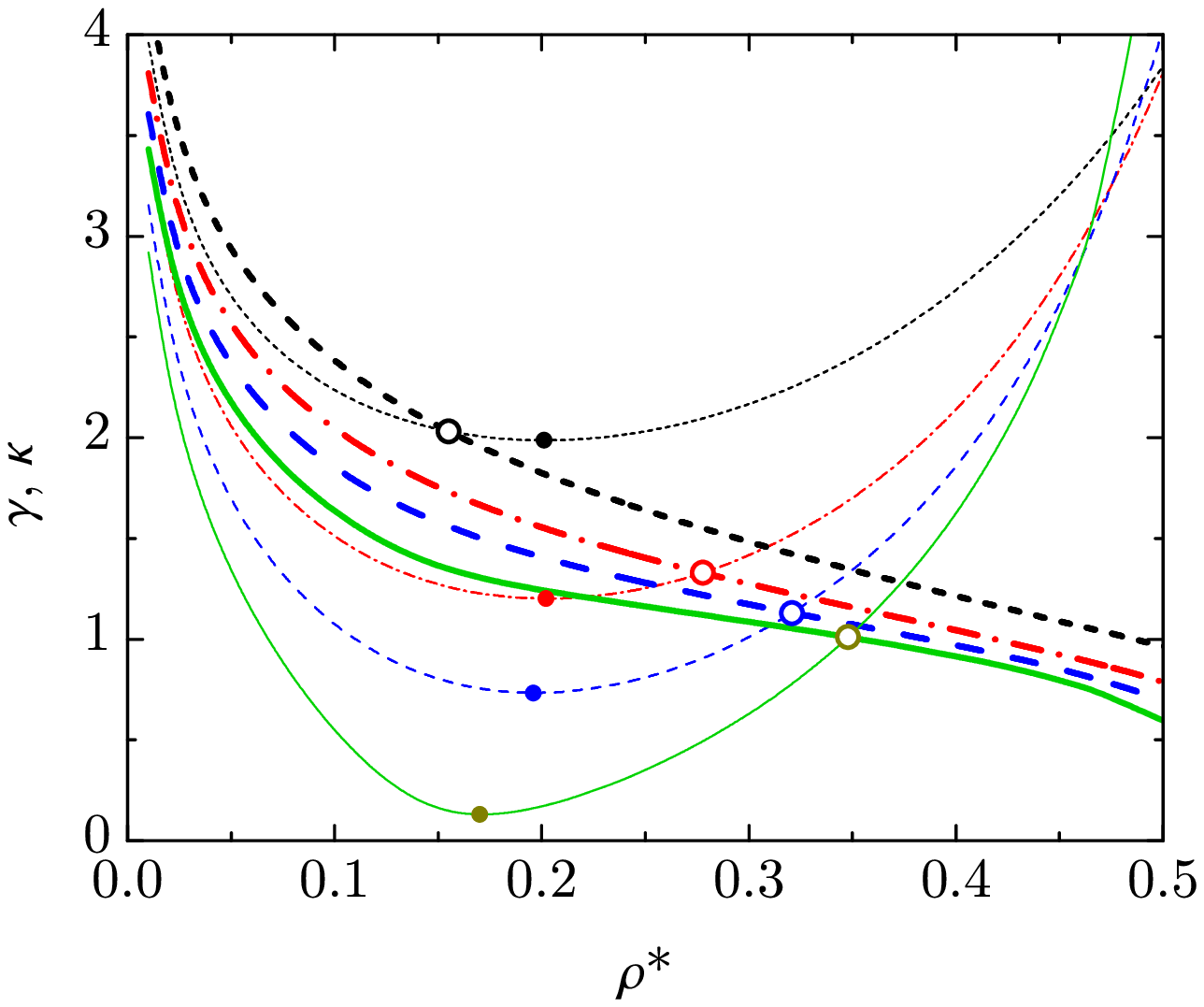}
\caption{Density dependence of the damping coefficients $\gamma$ (thick lines) and $\kappa$ (thin lines) in a fluid with a Jagla potential ($\lambda_1=1.3$, $\lambda_2=1.6$, $\varepsilon_1/\varepsilon_2=1$), as predicted by the RFA ($n=10$). The temperatures are $T^*=1.2$ (- - -), $T^*=0.8$ (-- $\cdot$ -- $\cdot$), $T^*={2}/{3}$ (-- -- --), and $T^*=0.58$ (---). At each temperature, the open circle marks the intersection point $\gamma=\kappa$ (FW transition point). The minima of the curves for $\kappa$ (solid circles) define the Widom line.
\label{fig:Poles}}
\end{figure}

\begin{figure}
\includegraphics[width=.9\columnwidth]{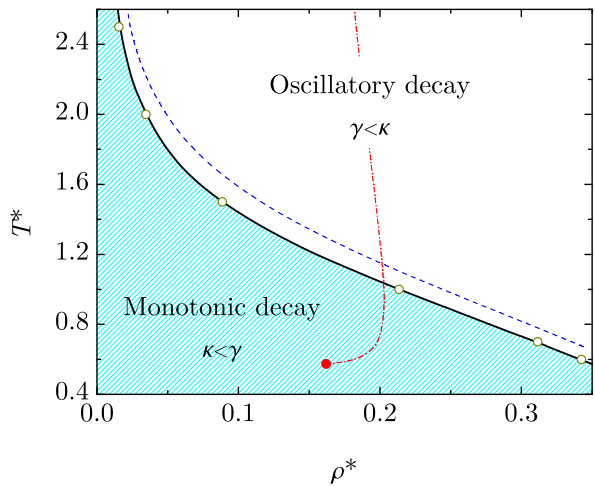}
\caption{Phase diagram in the $T^*$-$\rho^*$ plane for a fluid with a Jagla potential ($\lambda_1=1.3$, $\lambda_2=1.6$, $\varepsilon_1/\varepsilon_2=1$), as predicted by the RFA ($n=10$). The FW line (solid curve) splits the diagram into the hatched region where the decay of $h(r)$ is monotonic (i.e., $\kappa<\gamma$) and the region where the decay is oscillatory (i.e., $\gamma\leq\kappa$). The dash-dotted curve represents the Widom line, where, at a given temperature, $\kappa$ presents a minimum value. The Widom line inside the monotonic decay region terminates at the critical point (solid circle), where $\kappa\to 0$. The dashed curve is the FW line obtained from the  RFA  with $n=5$, while the open circles are points of the FW line obtained from the RFA with $n=20$.
\label{fig:FW}}
\end{figure}

\begin{figure}
\includegraphics[width=.9\columnwidth]{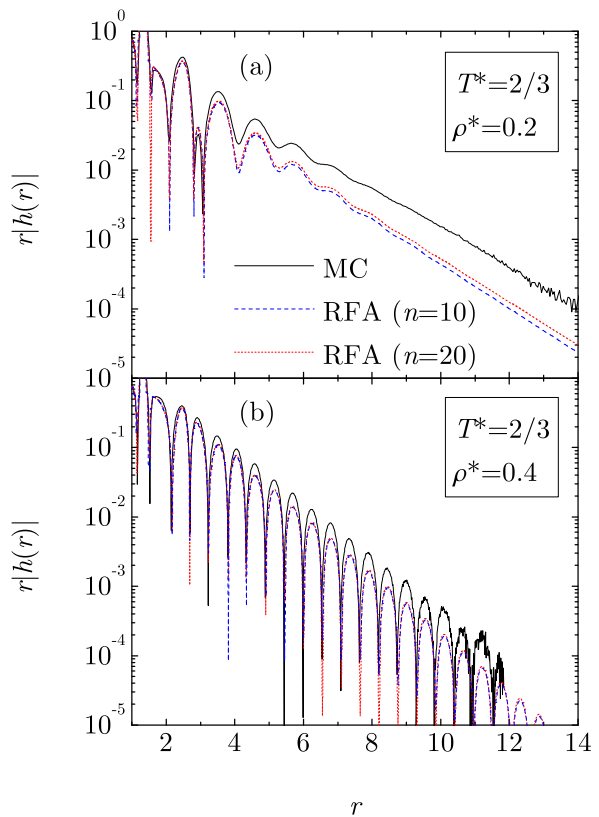}
\caption{Semilogarithmic plot of $r|h(r)|$  for a fluid with a Jagla potential ($\lambda_1=1.3$, $\lambda_2=1.6$, $\varepsilon_1/\varepsilon_2=1$) at (a) $(T^*,\rho^*)=({2}/{3},0.2)$ and (b) $(T^*,\rho^*)=({2}/{3},0.4)$. The solid, dashed, and dotted lines correspond to MC simulations, and to the RFA with $n=10$ and $n=20$, respectively.
\label{fig:rh}}
\end{figure}

\begin{figure}
\includegraphics[width=0.9\columnwidth]{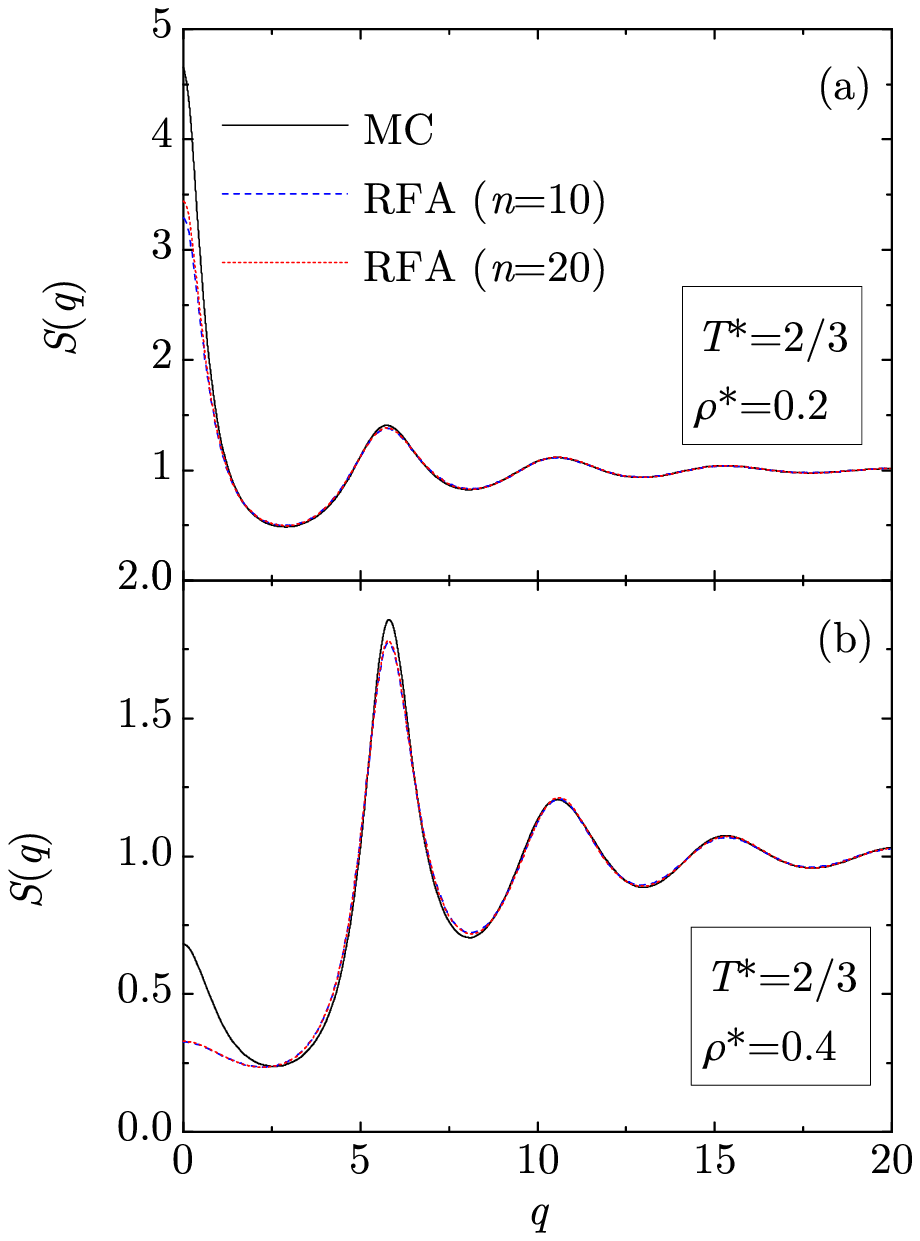}
\caption{
Structure factor of a fluid with a Jagla potential ($\lambda_1=1.3$, $\lambda_2=1.6$, $\varepsilon_1/\varepsilon_2=1$) at (a) $(T^*,\rho^*)=({2}/{3},0.2)$ and (b) $(T^*,\rho^*)=({2}/{3},0.4)$. The solid, dashed, and dotted lines correspond to MC simulations, and to the RFA with $n=10$ and $n=20$, respectively.
\label{fig:S_q}}
\end{figure}

According to general arguments \cite{DE00,FW69,PBH17,EHHPS93},  the total correlation function  $h(r)\equiv g(r)-1$ can be written as
\beq
\label{sec:fw1}
r h(r)= \sum_{i} \mathcal{A}_{i} e^{s_i r}
= \mathcal{A}_1 e^{s_1 r}+\mathcal{A}_2 e^{s_2 r}+\mathcal{A}_3 e^{s_3 r}+\cdots,
\eeq
where the sum runs over the discrete set of nonzero poles $s_i$ of the Laplace transform ${G}(s)$ of $rg(r)$, the ordering $0>\mathrm{Re}(s_1)\geq \mathrm{Re}(s_2)\geq \mathrm{Re}(s_3)\geq \cdots$ is adopted, and the amplitudes $\mathcal{A}_i=\mathrm{Res}\left[{G}(s)\right]_{s_i}$ are the associated (in general complex) residues. The \emph{asymptotic} decay  of  $h(r)$
is determined by the nature of the pole(s)  with the largest real part.

In general, in the case of potentials with an attractive part, the three dominating terms in Eq.\ \eqref{sec:fw1} are those associated with a pair of complex conjugate poles $s_{1,2}=-\gamma\pm\imath \omega$ (or $s_{2,3}=-\gamma\pm\imath \omega$), where $\imath$ denotes the imaginary unit, and a real pole $s_3=-\kappa$ (or $s_1=-\kappa$). Thus, the dominant behavior of $h(r)$ at large $r$ is
\begin{equation}
\label{asympt}
h(r)\approx \frac{1}{r}
\begin{cases}
2 |\mathcal{A}_\gamma| {e^{-\gamma r}}\cos(\omega r+\psi),& \gamma<\kappa,\\
\mathcal{A}_\kappa {e^{-\kappa r}},&\gamma>\kappa,
\end{cases}
\end{equation}
where $\psi$ is the argument of the residue $\mathcal{A}_{\gamma}$, i.e., $\mathcal{A}_{\gamma}=|\mathcal{A}_{\gamma}|e^{\pm \imath \psi}$.
Thus, the asymptotic behavior of $h(r)$ results from the competition between the real parts closest to the origin of the poles of $G(s)$. If $\gamma<\kappa$, a pair of conjugate complex poles dominate and the decay of the total correlation function is \emph{oscillatory}. On the other hand, if $\kappa<\gamma$, a real pole  is the dominant one and then the asymptotic decay is \emph{monotonic}. In the latter case, $\xi=\kappa^{-1}$ represents the \emph{correlation length}.

The oscillatory decay reflects the effects of the repulsive part of the interaction potential on spatial pair correlations, while the effects of the attractive part are reflected by the monotonic decay. Thus, at a given temperature, the first type of decay takes place at sufficiently high values of density, whereas the monotonic decay occurs at sufficiently low values of density. Following Fisher and Widom \cite{FW69}, the locus of transition points from one type to the other one ($\gamma=\kappa$) defines a line, the so-called {Fisher--Widom (FW) line}, in the temperature-versus-density plane. This line has been the subject of many investigations for different fluid models \cite{EHHPS93,VRL95,B96,DE00,TCV03}.

Since the RFA works, by construction, in Laplace space [see Eq.\ \eqref{b5}] and provides an \emph{explicit} dependence of $G(s)$ on $s$, it is ideally suited to determine the poles with a real part closest to the origin. As a representative example, we consider the Jagla fluid with $\lambda_1=1.3$, $\lambda_2=1.6$, and $\varepsilon_1/\varepsilon_2=1$. Moreover, the RFA with $n=10$ is generally employed.

Figure \ref{fig:Poles} shows the density dependence of both $\gamma$ and $\kappa$ at several temperatures. While, at a given temperature $T^*$, $\gamma$ monotonically decreases with increasing density, $\kappa$ presents a nonmonotonic behavior. As a consequence, both curves cross at a certain point of density $\rho^*_\text{FW}(T^*)$, such that $\gamma\leq\kappa$ (oscillatory decay) if $\rho^*\geq\rho^*_\text{FW}(T^*)$ and $\kappa<\gamma$ (monotonic decay) if $\rho^*<\rho^*_\text{FW}(T^*)$.

The locus of points $\rho^*_\text{FW}(T^*)$ defines the FW line, which is plotted in Fig.\ \ref{fig:FW}. The FW line predicted by the RFA with $n=5$, as well as a few points obtained with $n=20$, are also shown in Fig.\ \ref{fig:FW}. Although qualitatively analogous, the line with $n=5$ differs from that with $n=10$. However, moving from $n=10$ to $n=20$ has practically no effect on the FW line. Since $y(r)=g(r)$ for $r>\lambda_2$, the excellent agreement between the FW lines with $n=10$ and $n=20$ is a fine-grained test of the ansatz \eqref{yJagla}.

The nonmonotonic behavior of $\kappa$ versus $\rho^*$ at fixed $T^*$ observed in Fig.\ \ref{fig:Poles} implies the occurrence of a minimum value at a certain density $\rho^*_\text{W}(T^*)$. The locus $\rho^*_\text{W}(T^*)$ is also plotted in Fig.\ \ref{fig:FW}. Although it extends to the region of oscillatory decay, the line $\rho^*_\text{W}(T^*)$ is relevant only in the region of monotonic decay, i.e., below the FW line, where the asymptotic behavior is $h(r)\sim r^{-1} {e^{-\kappa r}}$ [see Eq.\ \eqref{asympt}]. In that region, the line $\rho^*_\text{W}(T^*)$ marks the states where the correlation length $\xi=\kappa^{-1}$ presents a maximum at a given temperature, so that it can be termed a \emph{Widom line}. In general, a Widom line refers to a locus of maximum response that ends at a critical point \cite{XKBCPSS05,XEBS06,XBAS06,BMAGPSSX09,LXASB15,RDMS17,RBMI17}. Thus, it represents an extension of the coexistence line into the one-phase region.
As shown in Fig.\ \ref{fig:Poles}, the Widom point at $T^*=0.58$ corresponds to a value of $\kappa$ rather close to zero, so $T^*=0.58$ is only slightly above the critical temperature $T_c^*$ (where $\kappa\to 0$). A more precise estimate of the critical point yields $T_c^*=0.574$ and $\rho_c^*=0.162$.

In order to assess the reliability of the RFA prediction for the FW line, we  performed detailed MC simulations of the Jagla fluid (again with $\lambda_1=1.3$, $\lambda_2=1.6$, and $\varepsilon_1/\varepsilon_2=1$). While a pole analysis similar to the one performed above can also be carried out using simulation data as input \cite{DE00}, here we focus on the direct measurement of $h(r)$ for large $r$. In this instance, we took $N = 5324$ particles and $2000$ independent simulations for each physical state, divided into four blocks of $500$ simulations each. In the first block, the system was aged for $10^5$ MC steps to reach equilibration and then $2 \times 10^5$ additional MC steps were performed for data collection. The final equilibrated states of the first block were taken as the initial states for the second block, and so on. Along the equilibrated
$2 \times 10^5$ MC steps for each simulation,  data were recorded every $50$ MC steps and averaged every $1000$ saved configurations. The averaged function $g(r)$ was evaluated  with a bin size $\delta r=0.01$.

We chose two representative states: (A) $(T^*,\rho^*)=({2}/{3},0.2)$ and (B) $(T^*,\rho^*)=({2}/{3},0.4)$.
According to Figs.\ \ref{fig:Poles} and \ref{fig:FW}, the RFA predicts that state A lies in the monotonic-decay region, while state B lies in the oscillatory-decay region. These predictions are confirmed by Fig.\ \ref{fig:rh}, which shows  $r|h(r)|$ as measured in our MC simulations and as obtained from the RFA (with $n=10$ and $n=20$). It must be noted that, due to unavoidable finite-size effects, the asymptotic value $g_\infty$ of $g(r)$ in the MC simulations is slightly smaller than $1$,  and this needs to be taken into account in the MC evaluation of $h(r)$ as $h(r)=g(r)-g_\infty$. In particular, $g_\infty=0.99912(1)$ and $g_\infty=0.99987(1)$ at states A and B, respectively. Therefore, the error in $h(r)$ is of order $10^{-5}$ and this is why the maximum accessible distance is  $r\sim 10$, which corresponds to $r|h(r)|\sim 10^{-4}$.

Figure \ref{fig:rh} not only confirms that the decay at states A and B is monotonic and oscillatory, respectively, but also that the RFA produces reasonable estimates of the damping coefficients  $\gamma$ and  $\kappa$, respectively. While the true values of $\gamma$ and  $\kappa$ are somewhat smaller than the RFA values, Fig.\ \ref{fig:rh}(b) shows an excellent theoretical prediction of the wavelength $2\pi/\omega$.
Moreover, a very good agreement exists between the results of the RFA approach and the MC simulation data for the region  prior to the asymptotic regime ($r\lesssim 4$).

Next, we turn to another issue related to the asymptotic behavior of $h(r)$ for the Jagla fluid. The (static) structure factor $S(q)$ and the total correlation function $h(r)$ are related by
\cite{HM06,S16}
\beq
S(q)=1+\rho\widetilde{h}(q),
\label{S(q)}
\eeq
where
\bal
\widetilde{h}(q)=&\int d\mathbf{r}\, e^{-\imath\mathbf{q}\cdot\mathbf{r}}h(r)\nn
=&\frac{4\pi}{q}\int_0^\infty dr\, r\sin(qr) h(r)
\label{h(q)}
\eal
is the Fourier transform of the total correlation function.  Thus, it is easy to check the following exact relationship between $S(q)$ and $G(s)$  \cite{S16},
\beq
S(q)=1-12\eta \left[\frac{G(s)-G(-s)}{s}\right]_{s=\imath q}.
\label{S2(q)}
\eeq

Since $S(0)$ coincides with the (reduced) isothermal compressibility, the low-$q$ behavior of $S(q)$, which is deeply related to the asymptotic behavior of $h(r)$ at large $r$, provides us with another source for performing the evaluation of the RFA approach regarding thermodynamic properties. The structure factors for the same two cases A and B examined in Fig.\ \ref{fig:rh}  are shown in Fig.\ \ref{fig:S_q}. The RFA curves were obtained from the analytic function $G^{(n)}(s)$ by application of Eq.\ \eqref{S2(q)}, so they actually represent the discretized potential $\phi^{(n)}(r)$. However, we have checked that the ``refined'' structure factor obtained by making use of Eq.\ \eqref{gJagla} is practically indistinguishable from the one obtained from $G^{(n)}(s)$. As for the MC curves, they were computed by a numerical  integration using the MC values of $h(r)$ from $r=0$ up to a cutoff distance $r=r_c\approx 12$,   plus the analytic integration from $r_c$ to infinity using the expressions of $h(r)$ at large $r$ given by Eq.\ \eqref{asympt}, with  a fit of $A_\kappa$ and $\kappa$ in case A, and of $|A_\gamma|$, $\gamma$, $\omega$, and $\psi$ in case B.

In agreement with what  is observed in Fig.\ \ref{fig:rh}, the RFA structure factors with $n=10$ and $n=20$ are practically indistinguishable from each other. Only a small increase of $S(0)$ (of about $3\%$) can be seen in Fig.\ \ref{fig:S_q}(a) when going from $n=10$ to $n=20$, thus reflecting a corresponding increase of the correlation length from $\xi=\kappa^{-1}=1.364$ to $\xi=\kappa^{-1}=1.398$. Regarding the comparison with the MC curves, a very good general agreement exists, except in the region $0\leq q\lesssim 2$. In that region,
for the two cases A and B considered in Figs.\ \ref{fig:rh} and \ref{fig:S_q},
$S(q)$ is  rather sensitive to the asymptotic behavior of $h(r)$. On the other hand, although not shown, it is important to point out here that the agreement between the results of the RFA approach and the MC simulation data for the structure factor $S(q)$ is very good, even near $q=0$, for states (such as those considered in Fig.\ \ref{fig:Jagla}) with a relatively rapid decay of $h(r)$.

%%%%%%%%%%%%%%%%%%%%%%%%%%%%%%%%%%%%%%%%%%%%%%%%%%%%%%%%%%%%%%%%%%%%%%%%%

\section{Concluding remarks}
\label{sec:concl}

{The results of the previous sections deserve further consideration. By replacing the actual Jagla potential by a discretized version consisting of a hard-core and a suitable piecewise constant function, and assuming a common cavity function, we have been able to compute a semi-analytical RFA for the radial distribution function of the Jagla fluid, and also for the two limit cases of the triangle-well fluid and the ramp fluid. In the illustrative examples that we have presented, this leads to a highly satisfactory agreement between the results of the RFA approach and the MC data. Such agreement is in some instances superior to the one exhibited by the results of the PY and HNC integral equation approximations. Although not shown, we have found that the above performance of the RFA improves when the potential range $\lambda_2$ decreases. On the other hand, it becomes poorer as either the temperature decreases or the density increases. Since the number $n$ of steps involved in the discretization leads to feasible numerical solutions of Eq.\ (\ref{c15_c}),  the method represents an excellent compromise between accuracy and simplicity. This is reinforced by the fact that going from $n = 10$ to $n = 20$ steps does not produce a significant difference in the numerical values of the radial distribution function.

The analysis of the asymptotic behavior of the total correlation function for large distances, as represented by the Fisher--Widom and  Widom lines, also confirms that the RFA approach for the Jagla fluid is both simple and useful. In fact, it produces very reasonable estimates of the damping coefficients for either the monotonic or oscillatory behavior and, in this latter instance, it  even leads to an excellent theoretical prediction of the wavelength. All of the above  provides support to the idea that a similar approach to the one followed here for the Jagla fluid may be profitably employed to compute the structural properties of fluids whose molecules interact with other continuous potentials.}

\begin{acknowledgments}
We are grateful to R.\ Evans for proposing us the study of the Fisher--Widom line. Financial support of the Spanish Agencia Estatal de Investigaci\'on through Grant No.\ FIS2016-76359-P and the Junta de Extremadura
(Spain) through Grant No.\ GR18079, both partially financed by FEDER funds, is acknowledged. M.L.H. thanks Universidad de Extremadura, where most of this work was carried out, and CONACYT (Mexico) for a sabbatical grant that made possible his stay there.
\end{acknowledgments}

\bibliography{D:/Dropbox/Mis_Dropcumentos/bib_files/liquid}

\end{document}